\documentclass[a4paper,11pt]{article}

\usepackage{amsmath}
\usepackage{amssymb}
\usepackage[dvips]{graphicx}
\usepackage[dvips]{psfrag}
\usepackage{cite}

\makeatletter
\@addtoreset{equation}{section}
\renewcommand{\theequation}{\thesection.\@arabic\c@equation}
\makeatother

\makeatletter
\renewcommand\appendix{\par
  \setcounter{section}{0}%
  \setcounter{subsection}{0}%
  \gdef\thesection{Appendix \@Alph\c@section }
  \renewcommand{\theequation}
  {\Alph{section}.\arabic{equation}}
}
\makeatother

\def \be {\begin{equation}}
\def \ee {\end{equation}}
\def \ba {\begin{array}}
\def \ea {\end{array}}
\def \bea{\begin{eqnarray}}
\def \eea{\end{eqnarray}}

\def \a {\alpha}
\def \b {\beta}
\def \g {\gamma}
\def \G {\Gamma}
\def \d {\delta}
\def \D {\Delta}

\def \m {\mu}

\def \k {\kappa}
\def \l {\lambda}
\def \L {\Lambda}
\def \s {\sigma}
\def \S {\Sigma}
\def \r {\rho}
\def \o {\omega}
\def \O {\Omega}
\def \th {\theta}

\def \p {\partial}
\def \f {\frac}
\def \na {\nabla}

\def \nn {\nonumber}
\def \ma {\mathcal}
\def \mb {\mathbb}

\def \lt {\left}
\def \rt {\right}

\def \ra {\rightarrow}
\def \sr {\sqrt}
\def \td {\tilde}
\def \hs {\hspace}
\def \pp {\propto}

\begin{document}

\titlepage

\vspace*{-15mm}
\baselineskip 10pt

\baselineskip 24pt
\vglue 10mm

\begin{center}
{{\Large\bf General Hidden Conformal Symmetry of 4D Kerr-Newman and 5D Kerr Black Holes}
}

\vspace{8mm}

\baselineskip 18pt

\renewcommand{\thefootnote}{\fnsymbol{footnote}}

Bin Chen$^{1,2}$\footnote[1]{\,bchen01@pku.edu.cn}
and
Jia-ju Zhang$^1$\footnote[2]{\,jjzhang@pku.edu.cn}

\renewcommand{\thefootnote}{\arabic{footnote}}

\vspace{5mm}

{\it
$^{1}$Department of Physics, and State Key Laboratory of Nuclear Physics and Technology,\\
Peking University, Beijing 100871, P.R. China\\
\vspace{2mm}
$^{2}$Center for High Energy Physics, Peking University, Beijing 100871, P.R. China
}

\vspace{10mm}

\end{center}

\begin{abstract}

There are two known CFT duals, namely the J-picture and the Q-picture, for a four-dimensional Kerr-Newman black hole, corresponding to the angular momentum $J$ and the electric charge $Q$ respectively. In our recent study we found a set of novel CFT duals for extremal Kerr-Newman black hole, incorporating these two pictures. The novel CFT duals are generated by the modular group $SL(2,\mb Z)$. In this paper we study these novel CFT duals for the generic non-extremal Kerr-Newman black hole. We investigate the hidden conformal symmetry in the low frequency scattering off Kerr-Newman black hole, from which the dual temperatures could be read. We find that there still exists a hidden conformal symmetry for a general CFT dual. We reproduce the correct Bekenstein-Hawking entropy from the Cardy formula, assuming the form of the central charge being invariant. Moreover we compute the retarded Green's function in the general CFT dual picture and find it  is in good match  with the CFT prediction. Furthermore we discuss the hidden conformal symmetries of the five dimensional Kerr black hole and obtain the similar evidence to support the general dual CFT pictures.

\end{abstract}

\baselineskip 18pt

\newpage


\section{Introduction}

In Kerr/CFT correspondence \cite{Guica:2008mu},  it was stated that an extreme Kerr black hole could be holographically described by an 2D chiral conformal field theory (CFT) with the non-vanishing temperature $T_L=1/2\pi$ and the central charge $c_L=12J$, where $J$ is the angular momentum of Kerr black hole. Then the Bekenstein-Hawking entropy $S_{BH}=2\pi J$ could be reproduced by the Cardy formula
\be \label{e3} S_{CFT}=\f{\pi^2}{3}c_L T_L. \ee
Such correspondence was soon generalized to various extreme black holes.\footnote{See the nice review \cite{Bredberg:2011hp} for complete references.}

In particular, the extreme black holes with multiple $U(1)$ symmetries are especially interesting. For such cases, it turns out that for each $U(1)$ there is a dual holographic CFT description correspondingly \cite{Hartman:2008pb,{Lu:2008jk}}. For example, for a four-dimensional (4D) Kerr-Newman(-AdS-dS) black hole, in addition to the $J$-picture corresponding to the rotation $U(1)$ symmetry, there is another $Q$-picture corresponding to the $U(1)$ gauge symmetry, which could be changed into a geometric $U(1)$ isometry when the metric is uplifted to five dimensions. Moreover, novel CFT duals of extremal Kerr-Newman(-AdS-dS) were proposed very recently in \cite{Chen:2011wm}, with the help of the stretched horizon formalism developed in \cite{Carlip:2011ax}. The novel CFT duals are generated by the modular group $SL(2,\mb Z)$. Every novel CFT is defined with respect to a Killing symmetry of translation along an angular variable, which could be the linear combination of original two Killing symmetries. The phenomenon that a general Killing symmetry defining a dual CFT has been found in higher dimensional Kerr black holes with multiple $U(1)$ rotational symmetries as well. In general, for an extreme Kerr-Newman black holes in higher dimensions with $n$ U(1) symmetries, the novel CFT duals could be generated by the $SL(n,\mb Z)$ T-duality group.

The CFT dual to an extreme black hole is actually not chiral. It was pointed out that there is actually a right-moving sector with the same central charge \cite{Matsuo:2009sj,{Castro:2009jf}}. But the excitations are suppressed in the extreme limit. When one consider the scattering off the extreme black hole, the black hole becomes near-extremal, the right-moving sector is excited. In the near-horizon limit, the
modes of interest are the ones near the super-radiant bound. It was shown in \cite{Bredberg:2009pv,Hartman:2009nz,Cvetic:2009jn,arXiv:1001.3208,Becker:2010jj}
that the bulk scattering amplitudes
were in precise agreement with those in the CFT description whose form is
completely fixed by the conformal invariance. Furthermore, the Kerr/CFT correspondence was developed even for generic non-extremal case. In a remarkable
paper \cite{Castro:2010fd} it was argued that the existence
of conformal invariance in a near horizon geometry is not a
necessary condition, instead the existence of a local conformal
invariance in the solution space of the wave equation for the
propagating field is sufficient to ensure a dual CFT description. The observation
indicates that even though the near-horizon geometry of a generic Kerr black hole could
be far from the AdS or warped AdS spacetime, the local conformal symmetry on the solution space may still
allow us to associate a CFT description to a Kerr black hole. The periodic identification of angular variable in the conformal coordinates not only breaks the $SL(2, \mb R)$ symmetry to $U(1)$, it also allows us to read the temperatures of dual CFT. With the assumption that the central charge is still  $c_L=c_R=12J$, the non-chiral Cardy's formula
\be \label{e4} S_{CFT}=\f{\pi^2}{3}(c_L T_L +c_R T_R) \ee
reproduces the non-extremal Bekenstein-Hawking entropy.
Further support to this picture comes from the agreement of  the low frequency scalar
scattering amplitudes in the  near region with the CFT prediction. It turned out that this hidden conformal symmetry is an intrinsic properties of the black hole, not an artifact of the scalar equation of motion\cite{Chen:2010ik}. For an interesting study on how the conformal symmetry appears, see \cite{Cvetic:2011hp}. The hidden conformal symmetry of various black holes have been studied widely in the literature. The  hidden conformal symmetry of Kerr-Newman black hole was studied  in \cite{Chen:2010as,Chen:2010xu,Chen:2010bh,Chen:2010ywa}. The hidden conformal symmetry of 5D Kerr was studied in \cite{Krishnan:2010pv,{Setare:2010sy}}. For other study on the hidden conformal symmetry in the Kerr/CFT correspondence, see \cite{HiddenSymmetry}.

In this paper we would like to investigate if the novel CFT duals found in \cite{Chen:2011wm} still make sense for the generic non-extremal case. We focus mainly on four-dimensional (4D) Kerr-Newman black hole and five-dimensional (5D) Kerr black hole. Though both kinds of black holes have the same feature that there are two $U(1)$ hairs, they are physically very different. Nevertheless for both cases, we show that there exists general hidden conformal symmetry as well in the low frequency scattering off the black holes, which allows us to read the dual temperatures directly. With the similar assumption that the form of the central charges should be unchanged, we reproduce exactly the macroscopic Bekenstein-Hawking entropies of the black holes. We also show that the retarded scalar Green's functions are in perfect match with the CFT predictions. All these facts suggest that the novel CFT dual pictures apply to the generic non-extremal black holes without trouble.

The paper is organized as follows. In Section 2 we give a brief review of the general CFT duals we got in \cite{Chen:2011wm} for extreme 4D Kerr-Newman and 5D Kerr. In Section 3 we introduce the hidden conformal symmetry of both extremal and non-extremal black holes  and discuss its implications on real-time correlators.  In Section 4, we investigate the general hidden conformal symmetries of 4D Kerr-Newman. In Section 5, we turn to 5D Kerr. We end with the conclusions and discussions in section 6.

\section{General CFT duals of extreme black holes}

In this section we  give a brief review of the general CFT dual pictures for 4D extreme Kerr-Newman black hole and 5D extreme Kerr black hole. More detailed discussion could be found in \cite{Chen:2011wm}.

\subsection{Kerr-Newman black hole}
For a Kerr-Newman black hole with mass $M$, angular momentum $J=Ma$ and electric charge $Q$, its metric takes the ADM form
\be \label{j2}
ds^2=-\f{\r^2\D}{\S^2}dt^2+\f{\r^2}{\D}dr^2+\r^2 d\th^2
        +h_{\phi\phi}\lt( d\phi +N^\phi dt \rt)^2,
\ee
with
\bea
&&\r^2=r^2+a^2\cos^2\th,
~~~\D=r^2-2Mr+a^2+Q^2, \nn\\
&&\S^2=(r^2+a^2)^2-\D a^2\sin^2\th, \nn\\
&&h_{\phi\phi}=\f{\S^2\sin^2\th}{\r^2},
~~~N^\phi=-\f{(2Mr-Q^2)a}{\S^2}. \nn
\eea
The gauge field is
\be A=-\f{Q r}{\r^2}(dt-a\sin^2\th d\phi). \ee
There are two horizons located at
\be
r_{\pm}=M\pm\sr{M^2-a^2-Q^2}.
\ee
When $M^2=a^2+Q^2$ is satisfied, the two horizons merge at $r_+=r_-=M$ and give us an extremal black hole. The Bekenstein-Hawking entropy, Hawking temperature, angular velocity of the horizon and the electric potential are respectively
\bea \label{en}
&&S_{BH}=\pi (r_+^2+a^2),\nn\\
&&T_H=\f{r_+-r_-}{4\pi(r_+^2+a^2)},\nn\\
&&\O_H=\f{a}{r_+^2+a^2},\nn\\
&&\Phi_H=\f{Qr_+}{r_+^2+a^2}.
\eea

For a Kerr-Newman black hole, there are two parameters $J$ and $Q$ besides the mass $M$, to characterize the black hole. These two hairs correspond to two $U(1)$ symmetries of the background, one is the rotational symmetry $U(1)_J$ whose generator is $\p_\phi$ and the other is the $U(1)_Q$ gauge symmetry, which changes to a rotational symmetry $\p_\chi$ when the metric is uplifted to five dimension. It has been shown that for each $U(1)$, there is correspondingly a CFT dual, namely the J-picture and Q-picture. In the J-picture we have the central charge and temperature
\be \label{jp}
c_L^J=12J,~~~T_L^J=\f{r_+^2+a^2}{4\pi J},
\ee
and in the Q-picture,
\be \label{qp}
c_L^Q=6Q^3,~~~T_L^Q=\f{r_+^2+a^2}{2\pi Q^3}.
\ee
Moreover, there are more general $U(1)$ symmetries, which is just the translation in a general angular direction
\be
\lt(\ba{c} \phi' \\  \chi' \ea \rt)=
\lt( \ba{cc} \a & \b \\ \g & \d \ea \rt)
\lt(\ba{c} \phi \\  \chi \ea \rt),
\ee
with
\be \lt( \ba{cc} \a & \b \\ \g & \d \ea \rt) \in SL(2,\mb Z). \ee
Correspondingly, there are more general dual pictures, namely $\phi'$ picture and $\chi'$ picture, with
\bea \label{gp}
&& c_L^{\phi'}=6(2\a J+\b Q^3), ~~~ T_L^{\phi'}=\f{r_+^2+a^2}{2\pi(2\a J+\b Q^3)},  \nn\\
&& c_L^{\chi'}=6(2\g J+\d Q^3), ~~~ T_L^{\chi'}=\f{r_+^2+a^2}{2\pi(2\g J+\d Q^3)}.
\eea
When
\be \lt( \ba{cc} \a & \b \\ \g & \d \ea \rt)=\lt( \ba{cc} 1 & 0 \\ 0 & 1 \ea \rt), \ee
we recover the J-picture and Q-picture respectively. With Cardy's formula (\ref{e3}) all the pictures can correctly reproduce the Bekenstein-Hawking entropy
\be \label{e5}
S_{CFT}=\f{\pi^2}{3}c_LT_L=\pi (r_+^2+a^2)=S_{BH}.
\ee

For non-extremal Kerr-Newman black hole, the hidden conformal symmetries in the J-picture and Q-picture  were studied in \cite{Chen:2010xu,HiddenSymmetry} and \cite{Chen:2010ywa} respectively. In the J-picture we have the temperatures
\bea
T_L^J=\f{r_+^2+r_-^2+2a^2}{8\pi J},~~~T_R^J=\f{r_+^2-r_-^2}{8\pi J},
\eea
and in the Q-picture
\be
T_L^Q=\f{r_+^2+r_-^2+2a^2}{4\pi Q^3},~~~T_R^Q=\f{r_+^2-r_-^2}{4\pi Q^3}.
\ee
If we assume in the non-extremal case, the left and right central charges are the same and unchanged from the extremal case
\bea\label{cbk}
&& c_L^J=c_R^J=12J, \nn\\
&& c_L^Q=c_R^Q=6Q^3,\nn\\
&& c_L^{\phi'}=c_R^{\phi'}=6(2\a J+\b Q^3), \nn\\
&& c_L^{\chi'}=c_R^{\chi'}=6(2\g J+\d Q^3),
\eea
then with the Cardy formula (\ref{e4}) we find that both J- and Q-pictures can correctly reproduce the Bekenstein-Hawking entropy
\be
S_{CFT}=\f{\pi^2}{3}(c_L T_L+c_R T_R)=\pi (r_+^2+a^2)=S_{BH}.
\ee
In this paper, we will show that the general pictures whose central charges are (\ref{cbk})  still make sense for generic non-extremal black holes.

\subsection{5D Kerr black hole}

The 5D rotating  Kerr black hole solution was firstly obtained by Myers and Perry in \cite{Myers:1986un}.  Its metric  takes the form
\bea
&&ds^2=-\f{\D}{\r^2}(dt-a\sin^2\th d\phi-b\cos^2\th d\psi)^2+\f{\r^2}{\D}dr^2+\r^2d\th^2\nn\\
&&\phantom{ds^2=}+\f{\sin^2\th}{\r^2}[adt-(r^2+a^2)d\phi]^2
                 +\f{\cos^2\th}{\r^2}[bdt-(r^2+b^2)d\psi]^2 \nn\\
&&\phantom{ds^2=}+\f{1}{r^2\r^2}[abdt-b(r^2+a^2)\sin^2\th d\phi-a(r^2+b^2)\cos^2\th d\psi]^2,
\eea
where
\be
\D=\f{1}{r^2}(r^2+a^2)(r^2+b^2)-2M,~~~\r^2=r^2+a^2\cos^2\th+b^2\sin^2\th.
\ee
The physical mass and two angular momenta of the black hole are
\be \label{z18}
\ma M=\f{3\pi M}{4},~~~J_\phi=\f{\pi Ma}{2},~~~J_\psi=\f{\pi Mb}{2}.
\ee
The two solutions of the equation $r^2 \D=0$ give us the outer and inner horizons $r_\pm$,
\be
r_\pm^2=M-\f{a^2+b^2}{2} \pm\sr{(M-\f{a^2+b^2}{2})^2-a^2b^2}.
\ee
When $M=(a+b)^2/2$, the black hole becomes extremal with $r_+=r_-=\sr{ab}$.
The surface gravity at the outer and inner horizons are
\be
\k_\pm=\f{r_\pm(r_+^2-r_-^2)}{(r_\pm^2+a^2)(r_\pm^2+b^2)},
\ee
and the Hawking temperature is
\be
T_H=\f{\k_+}{2\pi}=\f{r_+(r_+^2-r_-^2)}{2\pi(r_+^2+a^2)(r_+^2+b^2)}.
\ee
The angular momenta  of the two azimuthes at the outer and inner horizons are
\be
\O^\phi_\pm=\f{a}{r_\pm^2+a^2},~~~\O^\psi_\pm=\f{b}{r_\pm^2+b^2}.
\ee
There are also $\O_{R,L}$ defined in \cite{Cvetic:1997uw} as
\bea
&&\O_R=\O^\phi_+  +\O^\psi_+=\f{r_+(a+b)(r_++r_-)}{(r_+^2+a^2)(r_+^2+b^2)}, \nn \\
&&\O_L=\O^\phi_+  -\O^\psi_+=\f{r_+(a-b)(r_+-r_-)}{(r_+^2+a^2)(r_+^2+b^2)}.
\eea
The Bekenstein-Hawking entropy of the black hole is
\be \label{z2}
S_{BH}=\f{\pi^2(r_+^2+a^2)(r_+^2+b^2)}{2r_+},
\ee
and in the extremal limit it becomes
\be \label{z1}
S_{BH}=\f{\pi^2}{2}\sr{ab}(a+b)^2.
\ee

It was calculated in \cite{Lu:2008jk} that the central charges and the temperatures of the two chiral CFT pictures, namely the $\phi$ picture and $\psi$ picture, dual to the extreme 5D Kerr are
\bea
&&c_L^\phi=\f{3\pi}{2}b(a+b)^2,~~~T_L^\phi=\f{\sr{ab}}{\pi b},\nn\\
&&c_L^\psi=\f{3\pi}{2}a(a+b)^2,~~~T_L^\psi=\f{\sr{ab}}{\pi a}.
\eea
Moreover, in \cite{Chen:2011wm} it was argued that there are novel general dual pictures  parameterized by
\be \lt( \ba{cc} \a & \b \\ \g & \d \ea \rt) \in SL(2,\mb Z), \ee
with
\bea
&& c_L^{\phi'}=\f{3\pi}{2}(\a b+\b a)(a+b)^2,~~~T_L^{\phi'}=\f{\sr{ab}}{\pi(\a b+\b a)}, \nn\\
&& c_L^{\psi'}=\f{3\pi}{2}(\g b+\d a)(a+b)^2,~~~T_L^{\psi'}=\f{\sr{ab}}{\pi(\g b+\d a)}.
\eea
With Cardy's formula for chiral CFT, all pictures can  reproduce the Bekenstein-Hawking (\ref{z1}) entropy correctly
\be
S_{CFT}=\f{\pi^2}{3}c_L T_L=S_{BH}.
\ee

It is more illuminating to rewrite the central charge of the 5D Kerr black hole in terms of  physical quantities. And to discuss the generic non-extremal black hole, we  assume the  left and right central charges are the same and their forms are unchanged from the extremal case. Then  the central charges of
three different pictures could be respectively
\bea \label{z9}
&&c_L^\phi=c_R^\phi=6J_\psi=\f{3\pi b(r_+^2+a^2)(r_+^2+b^2)}{2r_+^2},  \nn\\
&&c_L^\psi=c_R^\psi=6J_\phi=\f{3\pi a(r_+^2+a^2)(r_+^2+b^2)}{2r_+^2}, \nn\\
&&c_L^{\phi'}=c_R^{\phi'}=6(\a J_\psi+\b J_\phi)=\f{3\pi(\a b+\b a)(r_+^2+a^2)(r_+^2+b^2)}{2r_+^2}, \nn\\
&&c_L^{\psi'}=c_R^{\psi'}=6(\g J_\psi+\d J_\phi)=\f{3\pi(\g b+\d a)(r_+^2+a^2)(r_+^2+b^2)}{2r_+^2}.
\eea

According to \cite{Krishnan:2010pv,Setare:2010sy}, there are two different pictures  for a general non-extremal 5D Kerr black hole, namely the $\phi$ picture and the $\psi$ picture, corresponding to two
$U(1)$ isometries. In the $\phi$ picture the temperatures are
\be \label{z10}
T_L^\phi=\f{r_+ + r_-}{2\pi b},~~~T_R^\phi=\f{r_+ - r_-}{2\pi b}.
\ee
Then with the central charge (\ref{z9}) and temperatures (\ref{z10})   the Bekenstein-Hawking entropy (\ref{z2})  could be reproduced from microscopic point of view
\be
S^\phi_{CFT}=\f{\pi^2}{3}(c_L^\phi T_L^\phi+c_R^\phi T_R^\phi)=S_{BH}.
\ee
Similarly in the so called $\psi$ picture, all the results can be obtained by exchanging $a \leftrightarrow b$ in the results of $\phi$ picture.
The general pictures for non-extremal 5D Kerr black hole will be investigated in section 5.

\section{Hidden Conformal Symmetry}

In this section we give a general review of the hidden conformal symmetries of both extremal and non-extremal black holes. 

\subsection{Extreme case}

For an extreme black hole, the conformal coordinates cannot be the extremal limit of those for non-extremal black hole. They were introduced in  \cite{Chen:2010fr} as
\bea
&&\o^+=\frac{1}{2}\left(\a_1 t+\b_1 \phi-\frac{\g_1}{r-r_+}\right),  \nn\\
&&\o^-= \frac{1}{2}e^{2\pi T_L\phi+2n_L t},  \nn\\
&&y=\sqrt{\frac{\g_1}{2(r-r_+)}}e^{\pi T_L\phi+n_Lt},
\eea
with which  the vector fields could be  locally defined
\bea
&&H_1=i\p_+ \nn\\
&&H_0=i\left(\o^+\p_++\frac{1}{2}y\p_y\right) \nn\\
&&H_{-1}=i(\o^{+2}\p_++\o^+y\p_y-y^2\p_-)
\eea
and
\bea
&&\tilde H_1=i\p_- \nn\\
&&\tilde H_0=i\left(\o^-\p_-+\frac{1}{2}y\p_y\right) \nn\\
&&\tilde H_{-1}=i(\o^{-2}\p_-+\o^-y\p_y-y^2\p_+).
\eea
These vector fields obey the $SL(2,\mb R)$ Lie algebra
\be
[H_0, H_{\pm 1}]=\mp iH_{\pm 1},\hs{5ex} [H_{-1},H_1]=-2iH_0,
\ee
and similarly for $(\tilde H_0, \tilde H_{\pm 1})$. The quadratic Casimir is
\bea
&&\ma H^2=\tilde{\ma H}^2=-H_0^2+\frac{1}{2}(H_1 H_{-1}+H_{-1}H_1) \nn\\
&&\phantom{\ma H^2}=\frac{1}{4}(y^2\p^2_y-y\p_y)+y^2 \p_+\p_-.
\eea
In terms of $(t,r,\phi)$ coordinates, the Casimir becomes
\be \label{e25}
\ma H^2=\p_r(r-r_+)^2\p_r-\f{\g_1^2(2\pi T_L\p_t-2n_L\p_\phi)^2}{A_2^2(r-r_+)^2}
                 -\f{2\g_1(2\pi T_L\p_t-2n_L\p_\phi)(\b_1\p_t-\a_1\p_\phi)}{A_2^2(r-r_+)},
\ee
with $|A_2|=2\pi T_L\a_1-2n_L\b_1$ and the sign of $A_2$ is chosen to make $\b_1/A_2>0$. If the black hole has Killing symmetry along $t$ and $\phi$, then the scalar field can be expanded as $\Phi=e^{-i\o t+im\phi}R(r)$, and the equation $\ma H^2\Phi=K\Phi$  gives us the radial equation of motion
\bea \label{z7}
&&\p_r(r-r_+)^2\p_r R(r)+\f{\g_1^2(2\pi T_L\o+2n_L m)^2}{A_2^2(r-r_+)^2}R(r) \nn\\
&& +\f{2\g_1(2\pi T_L\o+2n_L m)(\b_1 \o+\a_1 m)}{A_2^2(r-r_+)}R(r)=K R(r),
\eea
where $K$ is a constant.

Introducing
\be
z=-\f{2i\g_1(2\pi T_L\o+2n_L m)}{A_2(r-r_+)},
\ee
we get the Whittaker equation
\be
R''(z)+\lt( -\f{1}{4}+\f{k}{z}+\f{\f{1}{4}-n^2}{z^2} \rt)R(z)=0,
\ee
where
\be k=\f{i(\b_1 \o+\a_1 m)}{A_2},~~~n^2=\f{1}{4}+K. \ee
This equation has the solution
\be
R(z)=C_1 R_+(z)+C_2R_-(z),
\ee
where $R_\pm(z)=e^{-\f{z}{2}}z^{\f{1}{2}\pm n}F(\f{1}{2}\pm n-k,1\pm2n,z)$ are two linearly independent solution. Near the horizon $r\ra r_+$, $z\ra \infty$, the Kummer function could be expanded asymptotically
\be
F(\a,\g,z)\sim\f{\G(\g)}{\G(\g-\a)}e^{-i\a\pi}z^{-\a}+\f{\G(\g)}{\G(\a)}e^z z^{\a-\g}.
\ee
As we need to impose purely ingoing boundary condition at the horizon, we have to require
\be
C_1=-\f{\G(1-2n)}{\G(\f{1}{2}-n-k)}C,~~~C_2=\f{\G(1+2n)}{\G(\f{1}{2}+n-k)}C
\ee
to cancel the outgoing modes, where $C$ is a constant.

When $r\ra\infty$, $z\ra0$, $F(\a,\g,z)\ra 1$, and the solution has asymptotic behavior
\be
R \sim C_2 r^{h_L-1}+ C_1 r^{-h_L},
\ee
where $h_L$ is the conformal weight of the scalar
\be
h_L=\f{1}{2}+n=\f{1}{2}+\sr{\f{1}{4}+K}.
\ee

The coefficient $k$ can be written as
\be
k=i\f{\td \o_L}{2\pi T_L},
\ee
where the $\td \o_L$ is composed of the CFT parameters: the frequency $\o_L$, the charge $q_L$ and the chemical potential $\m_L$
\be\label{ol}
\td \o_L=\o_L-q_L \m_L,
\ee
with
\be \label{z12}
\o_L=\f{2\pi\b_1T_L\o}{A_2},~~~q_L=m,~~~\m_L=-\f{2\pi\a_1T_L}{A_2}.
\ee
Then the retarded Green's function could be read directly \cite{ChenChu}
\be
G_R \sim \f{C_1}{C_2}
    \pp \f{\G(h_L-i\f{\td \o_L}{2\pi T_L})}{\G(1-h_L-i\f{\td \o_L}{2\pi T_L})}
    \pp \sin[\pi(h_L+i\f{\td \o_L}{2\pi T_L})]\G(h_L-i\f{\td \o_L}{2\pi T_L})\G(h_L+i\f{\td \o_L}{2\pi T_L}),
\ee
which agrees with the CFT Euclidean correlator
\be
G_E \sim T_L^{2h_L-1}e^{ i  \td \o_{L,E}/2T_L }
          \G \lt( h_L- \f{\td \o_{L,E}}{2\pi T_L} \rt)
          \G \lt( h_L+ \f{\td \o_{L,E}}{2\pi T_L} \rt),
\ee
with the Euclidean frequency
\be \td \o_{L,E}=\o_{L,E}-i q_L \m_L,~~~\o_{L,E}=i\o_L. \ee

The absorption cross section can be read from the retarded Green's function
\bea
&&\s \sim \textrm{Im}G_R   \pp \sinh \lt( \f{ \td \o_L}{2T_L} \rt)
                               \lt| \G ( h_L+i \f{\td \o_L}{2\pi T_L} )\rt|^2,
\eea
which agrees with the finite temperature absorption cross section for a 2D chiral CFT
\be
\s \sim T_L^{2h_L-1} \sinh\lt( \f{ \td \o_L}{2T_L} \rt)
                \lt| \G ( h_L+i \f{\td \o_L}{2\pi T_L} )\rt|^2.
\ee

\subsection{Non-extremal case}
For the non-extreme black hole, the conformal coordinates could be defined as \cite{Castro:2010fd}
\bea
&&\o^+=\sr{\f{r-r_+}{r-r_-}}e^{2\pi T_R\phi+2n_R t},  \nn\\
&&\o^-=\sr{\f{r-r_+}{r-r_-}}e^{2\pi T_L\phi+2n_L t},  \nn\\
&&y=\sr{\f{r_+-r_-}{r-r_-}}e^{\pi(T_R+T_L)\phi+(n_R+n_L )t},
\eea
with which  the vector fields could be  locally defined
\bea
&&H_1=i\p_+, \nn\\
&&H_0=i\left(\o^+\p_++\frac{1}{2}y\p_y\right), \nn\\
&&H_{-1}=i(\o^{+2}\p_++\o^+y\p_y-y^2\p_-),
\eea
and
\bea
&&\tilde H_1=i\p_- \nn\\
&&\tilde H_0=i\left(\o^-\p_-+\frac{1}{2}y\p_y\right) \nn\\
&&\tilde H_{-1}=i(\o^{-2}\p_-+\o^-y\p_y-y^2\p_+).
\eea
These vector fields obey the $SL(2,\mb R)$ Lie algebra
\be
[H_0, H_{\pm 1}]=\mp iH_{\pm 1},\hs{5ex} [H_{-1},H_1]=-2iH_0,
\ee
and similarly for $(\tilde H_0, \tilde H_{\pm 1})$.

The quadratic Casimir is
\bea
&&\ma H^2=\tilde{\ma H}^2=-H_0^2+\frac{1}{2}(H_1 H_{-1}+H_{-1}H_1) \nn\\
&&\phantom{\ma H^2}=\frac{1}{4}(y^2\p^2_y-y\p_y)+y^2 \p_+\p_-.
\eea
In terms of $(t,r,\phi)$ coordinates, the Casimir becomes
\bea \label{e26}
&&\ma H^2=\p_r(r-r_+)(r-r_-)\p_r
-\f{(r_+-r_-)[\pi(T_L+T_R)\p_t-(n_L+n_R)\p_\phi]^2}{16\pi^2 A_1^2 (r-r_+)} \nn\\
&&\phantom{\ma H^2=} +\f{(r_+-r_-)[\pi(T_L-T_R)\p_t-(n_L-n_R)\p_\phi]^2}{16\pi^2 A_1^2 (r-r_-)},
\eea
with $A_1=T_L n_R-T_R n_L$. With the scalar field being expanded as $\Phi=e^{-i\o t+im\phi}R(r)$, the equation $\ma H^2\Phi=K\Phi$  gives us the radial equation of motion
\bea \label{z3}
&&\p_r(r-r_+)(r-r_-)\p_r R(r)
+\f{(r_+-r_-)[\pi(T_L+T_R)\o+(n_L+n_R)m]^2}{16\pi^2 A_1^2 (r-r_+)} R(r) \nn\\
&&-\f{(r_+-r_-)[\pi(T_L-T_R)\o+(n_L-n_R)m]^2}{16\pi^2 A_1^2 (r-r_-)} R(r)=K R(r),
\eea
where $K$ is a constant.

The equation (\ref{z3}) can be solved in term of the new variable
\be
z=\f{r-r_+}{r-r_-},
\ee
and the solutions include ingoing and outgoing modes as
\bea
&&R^{(in)}=z^{-i\g}(1-z)^{h}F(a,b;c;z),\nn\\
&&R^{(out)}=z^{i\g}(1-z)^{h}F(a^*,b^*;c^*;z),
\eea
where $F(a,b;c;z)$ are the hypergeometric functions and
\bea
&&h=\f{1}{2}+\sr{\f{1}{4}+K},\nn\\
&&\g=\f{\pi(T_L+T_R)\o+(n_L+n_R)m}{4\pi |A_1|},\nn\\
&&a=h-i\f{\pi T_L \o+ n_L m}{2\pi |A_1|},\nn\\
&&b=h-i\f{\pi T_R \o+ n_R m}{2\pi |A_1|},\nn\\
&&c=1-i2\g.
\eea
As $r\ra\infty$,  $z\ra1,~1-z\ra r^{-1}$, and the ingoing modes behave asymptotically
\be R^{(in)} \sim A r^{h-1}+B r^{-h}, \ee
with
\be
A=\f{\G(2h-1)\G(c)}{\G(a)\G(b)},
~~~B=\f{\G(1-2h)\G(c)}{\G(c-a)\G(c-b)},
\ee
and the conformal weight being
\be h_L=h_R=h=\f{1}{2}+\sr{\f{1}{4}+K}. \ee
Hence, the coefficients $a,~b$ can be expressed in terms of conformal weights  and two parameters $(\td \o_L,\td \o_R)$
\be
a=h_R-i\f{\td \o_R}{2\pi T_R},~~~
b=h_L-i\f{\td \o_L}{2\pi T_L},
\ee
and so,
\be
\g=\f{\td \o_L}{4\pi T_L}+\f{\td \o_R}{4\pi T_R},
\ee
where $(\td \o_L,\td \o_R)$ are composed of three sets CFT parameters : the frequencies $(\o_L,\o_R)$, the charges $(q_L,q_R)$ and the chemical potentials $(\m_L,\m_R)$
\be
\td \o_L=\o_L-q_L\m_L,~~~
\td \o_R=\o_R-q_R\m_R,
\ee
with
\bea \label{z11}
&&\o_L=\o_R=\f{\pi T_L T_R \o}{|A_1|},  \nn\\
&&q_L=q_R=m,\nn\\
&&\m_L=-\f{T_L n_R}{|A_1|},~~~\m_R=-\f{T_R n_L}{|A_1|}.
\eea

The retarded Green's function can be read from $A,~B$
\bea
&&G_R \sim \f{B}{A} \pp \f{\G(a)\G(b)}{\G(c-a)\G(c-b)} \nn\\
&&  \phantom{G_R}  \pp \sin \lt(\pi h_L+ i \f{ \td \o_L}{2T_L} \rt)
                          \sin \lt(\pi h_R+ i \f{ \td \o_R}{2T_R} \rt)  \nn\\
&&\phantom{G_R\pp} \times \G \lt( h_L-i \f{\td \o_L}{2\pi T_L} \rt)
                            \G \lt( h_L+i \f{\td \o_L}{2\pi T_L} \rt)  \nn\\
&&\phantom{G_R\pp} \times\G \lt( h_R-i \f{\td \o_R}{2\pi T_R} \rt)
                            \G \lt( h_R+i \f{\td \o_R}{2\pi T_R} \rt).
\eea
This agrees with the CFT Euclidean correlator
\bea
&&G_E \sim T_L^{2h_L-1}T_R^{2h_R-1}
         e^{ i  \td \o_{L,E}/2T_L }e^{ i  \td \o_{R,E}/2T_R }\nn\\
&&\phantom{G_E \sim}        \times \G \lt( h_L- \f{\td \o_{L,E}}{2\pi T_L} \rt)
                                   \G \lt( h_L+ \f{\td \o_{L,E}}{2\pi T_L} \rt) \nn\\
&&\phantom{G_E \sim}        \times \G \lt( h_R- \f{\td \o_{R,E}}{2\pi T_R} \rt)
                                   \G \lt( h_R+ \f{\td \o_{R,E}}{2\pi T_R} \rt),
\eea
with the Euclidean frequencies
\bea
&& \td \o_{L,E}=\o_{L,E}-i q_L \m_L,~~~\o_{L,E}=i\o_L,\nn\\
&& \td \o_{R,E}=\o_{R,E}-i q_R \m_R,~~~\o_{R,E}=i\o_R.
\eea

The absorption cross section can be read from the retarded Green's function
\bea
&&\s \sim \textrm{Im}G_R  \pp  \sinh \lt( \f{\td \o_L}{2T_L}+ \f{\td \o_R}{2T_R}\rt)
                      \lt| \G \lt( h_L+i\f{\td \o_L}{2 \pi T_L} \rt) \rt|^2
                      \lt| \G \lt( h_R+i\f{\td \o_R}{2 \pi T_R} \rt) \rt|^2,
\eea
which agrees with the finite temperature absorption cross section for a 2D CFT
\bea
&&\s \sim T_L^{2h_L-1} T_R^{2h_R-1}
     \sinh\lt( \f{ \td \o_L}{2T_L} + \f{ \td \o_R}{2T_R} \rt) \nn\\
&&\phantom{\s \sim}   \times \lt| \G ( h_L+i \f{\td \o_L}{2\pi T_L} )\rt|^2
                             \lt| \G ( h_R+i \f{\td \o_R}{2\pi T_R} )\rt|^2.
\eea

\section{General pictures for 4D Kerr-Newman Black Hole}

In this section, we discuss the hidden conformal symmetries in the 4D Kerr-Newman black hole. Firstly we study the extreme case and re-derive the temperature in the general dual picture. Next we discuss the non-extreme case and find the hidden conformal symmetry in the low-frequency scattering, from which we identify the left and right temperatures. Then we successfully reproduce the macroscopic entropy via the Cardy formula. Moreover from the first law of black hole thermodynamics, we identify the CFT conjugate charges which allow us to rewrite the low frequency scattering amplitudes   in a form consistent with the CFT prediction.

\subsection{Charged massive scalar wave equation}

In a 4D Kerr-Newman black hole background, the scalar field can be expanded as $\Phi=e^{-i\o t+im\phi+ie \chi}R(r)S(\th)$, with $\chi$ being explained as the coordinate of the internal space of the $U(1)$ symmetry. Thus the Klein-Gordon equation for a charged massive scalar
\be
(\na_\m-ieA_\m)(\na^\m-ieA^\m)\Phi-\m^2\Phi=0
\ee
can be decomposed into the angular and the radial equation
\be
\f{1}{\sin\th}\p_\th\sin\th\p_\th S(\th)
     +\lt(\l-a^2(\o^2-\m^2)\sin^2\th-\f{m^2}{\sin^2\th}\rt)S(\th)=0,
\ee
\be
\p_r\D\p_r R(r)
    +\lt( \f{[(r^2+a^2)\o-Qer-am]^2}{\D}-\m^2(r^2+a^2)+2ma\o-\l \rt)R(r)=0,
\ee
where $\l$ is the separation constant.

For a non-extremal Kerr-Newman black hole the radial equation can be recast in the following form
\bea
&&\p_r(r-r_+)(r-r_-)\p_r R(r)+\f{[(r_+^2+a^2)\o-am-Qr_+e]^2}{(r-r_+)(r_+-r_-)}R(r) \nn\\
&&-\f{[(r_-^2+a^2)\o-am-Qr_-e]^2}{(r-r_-)(r_+-r_-)}R(r) \nn\\
&&+[(\o^2-\m^2) r^2+2(M\o-Q e)\o r+\o^2(a^2-Q^2)+(2\o M -Qe)^2]R(r)\nn\\
&&=(\l+\m^2 a^2) R(r).
\eea
The potential terms in the last line of left-handed side can be neglected if we make the following assumptions \cite{Chen:2010ywa}: (1)small frequency $\o M\ll1$ (consequently $\o a\ll1$ and $\o Q\ll1$), (2)small probe mass $\m M\ll1$ (consequently $\m a\ll1$ and $\m Q\ll1$), (3)small probe charge $Qe\ll1$, (4)near region $\td \o r\ll1$ with $\td \o=\max\{\o,\m\}$. Then we have the polar angular and radial equations
\be
\f{1}{\sin\th}\p_\th\sin\th\p_\th S(\th)+\lt(\l-\f{m^2}{\sin^2\th}\rt)S(\th)=0,
\ee
\bea \label{e1}
&&\p_r(r-r_+)(r-r_-)\p_r R(r)+\f{[(r_+^2+a^2)\o-am-Qr_+e]^2}{(r-r_+)(r_+-r_-)}R(r) \nn\\
&&-\f{[(r_-^2+a^2)\o-am-Qr_-e]^2}{(r-r_-)(r_+-r_-)}R(r)=K R(r),
\eea
where we have $\l=l(l+1)$ and $K=l(l+1)+\m^2 a^2$.

Similarly for an extreme Kerr-Newman black hole, the radial equation is
\bea \label{e19}
&&\p_r(r-r_+)^2\p_r R(r)+\f{[(r_+^2+a^2)\o-am-Qr_+e]^2}{(r-r_+)^2}R(r) \nn\\
&&+\f{2(2r_+\o-Qe)[(r_+^2+a^2)\o-am-Qr_+e]}{r-r_+}R(r)=K R(r).
\eea

\subsection{Extremal case}

Here we use the new angular coordinates $\phi'=\a\phi+\b\chi$ and $\chi'=\g\phi+\d\chi$,with
\be \lt( \ba{cc} \a & \b \\ \g & \d \ea \rt) \in SL(2,\mb Z). \ee
The identification $e^{im\phi+ie\chi}=e^{im'\phi'+ie'\chi'}$ gives us
\be
m=\a m'+\g e', ~~~ e=\b m'+\d e'.
\ee
In order to get the dual picture corresponding to the $\phi'$ direction, we have to turn off the momentum mode along the $\chi'$ direction  by imposing $e'=0$. Then (\ref{e19}) becomes
\bea \label{e24}
&&\p_r(r-r_+)^2\p_r R(r)+\f{[(r_+^2+a^2)\o-(\a a+\b Q r_+)m']^2}{(r-r_+)^2}R(r) \nn\\
&&+\f{2(2r_+\o-\b Q m')[(r_+^2+a^2)\o-(\a a+\b Q r_+)m']}{r-r_+}R(r) \nn\\
&&=K R(r).
\eea
After we make the change $\p_\phi \ra \p_{\phi'},~m \ra m'$ in (\ref{e25}) and (\ref{z7}), we find the agreement between (\ref{z7}) with (\ref{e24}) under the identification
\bea \label{tc}
&&\b_1^{\phi'}=-\f{2r_+}{\b Q}\a_1^{\phi'},~~~\g_1^{\phi'}=-\f{2\a J+\b Q^3}{\b Q}\a_1^{\phi'}, \nn\\
&&n_L^{\phi'}=-\f{\a a+\b Q r_+}{2(2\a J+\b Q^3)},
    ~~~T_L^{\phi'}=\f{r_+^2+a^2}{2\pi(2\a J+\b Q^3)}.
\eea
This agreement shows that there exist a hidden conformal symmetry in the low-frequency scattering off the Kerr-Newman
black hole. The identification (\ref{tc}) gives us the temperature of the dual CFT, in perfect agreement with  the temperature in the general picture (\ref{gp}).

Since the CFT dual to an extreme black hole has only left non-vanishing temperature, the microscopic entropy comes from only the left sector.  It is easy to see that Cardy's formula (\ref{e3})  reproduce exactly the  Bekenstein-Hawking entropy of extremal Kerr-Newman black hole
\be
S_{BH}=\pi(r_+^2+a^2)=\pi \lt( M^2+\f{J^2}{M^2} \rt).
\ee
Taking variation directly, we have
\be
\d S_{BH}=2\pi\lt[ \lt( M-\f{J^2}{M^3} \rt)\d M +\f{J}{M^2} \d J \rt].
\ee
With the constraint of the extremal condition $M^2=J^2/M^2+Q^2$ giving
\be
\lt( M+\f{J^2}{M^3} \rt)\d M-\f{J}{M^2} \d J-Q \d Q=0,
\ee
we have
\be \label{e6}
\d S_{BH}=2\pi(2M\d M-Q\d Q).
\ee

The CFT conjugate charge $\d E_L^{\phi'}$ is defined as
\be \label{e12}
\d S_{CFT}^{\phi'}=\f{\d E_L^{\phi'}}{T_L^{\phi'}},
\ee
and the identification with (\ref{e6})  gives us
\be
\d E_L^{\phi'}=\f{(2M^2-Q^2)(2M\d M-Q\d Q)}{2\a J+\b Q^3}.
\ee
The identifications of parameters are $\d M=\o,~\d J=m=\a m'+\g e'$ and $\d Q=e=\b m'+\d e'$. Since the probe scalar does not have the momentum mode along $\chi'$ direction, $e'=0$,  we have
\be
\td \o_L^{\phi'}=\d E_L^{\phi'} \lt( \d M=\o;\d J=\a m';\d Q=\b m' \rt),
\ee
which agrees with the relation (\ref{ol}), now with
\be
\o_L^{\phi'}=\f{(2M^2-Q^2)2M\o}{2\a J+ \b Q^3},~~~q_L^{\phi'}=m',
    ~~~\m_L^{\phi'}=\f{(2M^2-Q^2)\b Q}{2\a J+\b Q^3}.
\ee
Therefore we show that the first law of thermodynamics gives us the correct
identification of quantum numbers such that the scattering amplitude is in perfect match
with the CFT prediction.

Similarly, we can show that there are CFT dual corresponds to the $\chi'$ direction with the central charges and temperatures,
\be c_L^{\chi'}=6(2\g J+\d Q^3), ~~~ T_L^{\chi'}=\f{r_+^2+a^2}{2\pi(2\g J+\d Q^3)}. \ee

\subsection{Non-extremal case}

For the general picture we set $e'=0$, and (\ref{e1}) becomes
\bea \label{e2}
&&\p_r(r-r_+)(r-r_-)\p_r R(r)+\f{[(r_+^2+a^2)\o-(\a a+\b Qr_+)m']^2}{(r-r_+)(r_+-r_-)}R(r) \nn\\
&&-\f{[(r_-^2+a^2)\o-(\a a+\b Q r_-)m']^2}{(r-r_-)(r_+-r_-)}R(r) \nn\\
&&=K R(r).
\eea
To find the agreement with the quadratic $SL(2,\mb R)$ Casimir equation (\ref{e26}),  we have
\bea
&&n_L^{\phi'}=-\f{2\a a+\b Q(r_++r_-)}{4(2\a J+\b Q^3)},
~~~n_R^{\phi'}=-\f{\b Q(r_+-r_-)}{4(2\a J+\b Q^3)},\nn\\
&&T_L^{\phi'}=\f{r_+^2+r_-^2+2a^2}{4\pi(2\a J+\b Q^3)},
~~~T_R^{\phi'}=\f{r_+^2-r_-^2}{4\pi(2\a J+\b Q^3)}.
\eea
The central charge are supposed to be
\be
c_L^{\phi'}=c_R^{\phi'}=6(2\a J+\b Q^3),
\ee
and then the CFT microscopic entropy gives precisely the Bekenstein-Hawking entropy
\be
S^{\phi'}_{CFT}=\f{\pi^2}{3}(c_L^{\phi'} T_L^{\phi'}+c_R^{\phi'} T_R^{\phi'})=\pi(r_+^2+a^2)=S_{BH}.
\ee

The three sets CFT parameters, namely the frequencies $(\o^{\phi'}_L,\o^{\phi'}_R)$, the charges $(q_L^{\phi'},q_R^{\phi'})$ and the chemical potentials $(\m_L^{\phi'},\m_R^{\phi'})$, may combine into two quantities ($\td \o^{\phi'}_L,\td \o^{\phi'}_R$) appearing in the scattering amplitude
\be\label{olr}
\td \o^{\phi'}_L=\o^{\phi'}_L-q_L^{\phi'}\m_L^{\phi'},~~~
\td \o^{\phi'}_R=\o^{\phi'}_R-q_R^{\phi'}\m_R^{\phi'},
\ee
with
\bea
&&\o_L^{\phi'}=\o_R^{\phi'}=\f{\o(r_++r_-)(r_+^2+r_-^2+2a^2)}{2(2\a J+\b Q^3)}\nn\\
&&q_L^{\phi'}=q_R^{\phi'}=m',\nn\\
&&\m_L^{\phi'}=\f{\b Q(r_+^2+r_-^2+2a^2)}{2(2\a J+\b Q^3)},\nn\\
&&\m_R^{\phi'}=\f{[2\a a+\b Q (r_++r_-)](r_++r_-)}{2(2\a J+\b Q^3)}.
\eea

On the other hand, from the first law of black hole thermodynamics
\be
\d S_{BH}=\f{\d M-\O_H \d J-\Phi_H \d Q}{T_H}
\ee
and its CFT counter-terms
\be
\d S_{CFT}^{\phi'}=\f{\d E_L^{\phi'}}{T_L^{\phi'}}+\f{\d E_R^{\phi'}}{T_R^{\phi'}},
\ee
we get the conjugate charges $(\d E_L^{\phi'},\d E_R^{\phi'})$ as
\bea
&&\d E_L^{\phi'}=\f{(2M^2-Q^2)(2M\d M-Q\d Q)}{2\a J+\b Q^3}, \nn\\
&&\d E_R^{\phi'}=\f{(2M^2-Q^2)2M\d M -2J\d J -2M^2Q\d Q}{2\a J+\b Q^3}.
\eea
The identifications of parameters are $\d M=\o,~\d J=m=\a m'+\g e'$ and $\d Q=e=\b m'+\d e'$. Since the probe scalar does not have the momentum mode along $\chi'$ direction, $e'=0$, we have
\bea
&&\td \o_L^{\phi'}=\d E_L^{\phi'} \lt( \d M=\o;\d J=\a m';\d Q=\b m' \rt), \nn\\
&&\td \o_R^{\phi'}=\d E_R^{\phi'} \lt( \d M=\o;\d J=\a m';\d Q=\b m' \rt),
\eea
which agree precisely with the relation (\ref{olr}). In other words, the black hole thermodynamics gives the
correct identifications such that the scattering amplitudes are in good match with the CFT prediction.

Also, there are the $\chi'$-picture CFT dual with
\bea
&& c_L^{\chi'}=c_R^{\chi'}=6(2\g J+\d Q^3),  \nn\\
&& T_L^{\chi'}=\f{r_+^2+r_-^2+2a^2}{4\pi(2\g J+\d Q^3)},
~~~T_R^{\chi'}=\f{r_+^2-r_-^2}{4\pi(2\g J+\d Q^3)}.
\eea

\section{General pictures for 5D Kerr Black Hole}

Physically, the 5D Kerr black hole is completely different from 4D Kerr-Newman. It has two $U(1)$ rotational symmetries, defining two
conserved angular momenta. In \cite{Chen:2011wm}, we showed that there exist novel CFT duals for extreme 5D Kerr. In this section, we
try to study these CFT duals for generic non-extremal 5D Kerr. We start from the low frequency scalar scattering off the black hole and
investigate the existence of hidden conformal symmetry. In this case, we actually need to modify the conformal coordinates slightly. More precisely
we should replace the $r$ and $r_\pm$ in the conformal coordinates introduced in section 3 with $r^2$ and $r_\pm^2$. Then we find that for both extremal and non-extremal 5D Kerr black hole, there exist general hidden conformal symmetry.

\subsection{Scalar scattering}

The scattering of massless scalar off a 5D Kerr background has been thoroughly studied in \cite{Cvetic:1997uw}. In this background, the scalar field could be decomposed as
\be
\Phi=R(r)S(\th)e^{-i\o t+im_\phi\phi+im_\psi\psi}
    =R(r)S(\th)e^{-i\o t+im_R(\phi+\psi)+im_L(\phi-\psi)},
\ee
with $m_{\phi,\psi}=m_R \pm m_L$. After separation of variables, the radial equation turns out to be
\bea \label{z4}
&&\f{\p}{\p x}\lt(x^2-\f{1}{4}\rt)\f{\p}{\p x}R+\f{1}{4} \lt[x \D \o^2-\L+M\o^2
   +\f{1}{x-\f{1}{2}}\lt( \f{\o}{\k_+}-\f{m_R\O_R+m_L\O_L}{\k_+}   \rt)  \rt. \nn\\
&& \phantom{\f{\p}{\p x}\lt(x^2-\f{1}{4}\rt)\f{\p}{\p x}R+\f{1}{4} \lt[\rt.}
   \lt. -\f{1}{x+\f{1}{2}}\lt( \f{\o}{\k_-}-\f{m_R\O_R-m_L\O_L}{\k_+}   \rt)  \rt]R=0,
\eea
where $\D$ is a constant we will never need, $\L$ stands for the eigenvalue of the angular Laplacian, and $x$ is defined as
\be
x=\f{r^2-\f{1}{2}(r_+^2+r_-^2)}{r_+^2-r_-^2}.
\ee
Under the low frequency limit $M\o \ll 1$ and in the near region $r \ll \f{1}{\o}$, (\ref{z4}) is just
\bea \label{z5}
&&\p_{r^2}(r^2-r_+^2)(r^2-r_-^2)\p_{r^2}R(r)
   +\f{(r_+^2-r_-^2)[\o-(\O_+^\phi m_\phi+\O_+^\psi m_\psi)]^2}{4\k_+^2(r^2-r_+^2)}R(r) \nn\\
&& -\f{(r_+^2-r_-^2)[\o-(\O_-^\phi m_\phi+\O_-^\psi m_\psi)]^2}{4\k_-^2(r^2-r_-^2)}R(r)=K R(r),
\eea
where  $\L$ into $K=\L/4$.
In the extremal limit,  (\ref{z5}) can be recast into
\bea \label{z6}
&&\p_{r^2}(r^2-r_+^2)^2\p_{r^2}R(r)
    +\f{ab(a+b)^4}{4(r^2-r_+^2)^2}\lt(\o-\f{m_\phi+m_\psi}{a+b}\rt)^2 R(r)  \nn\\
&&+\f{(a+b)^4}{4(r^2-r_+^2)} \lt(\o-\f{m_\phi+m_\psi}{a+b}\rt)
                             \lt[ \o-\f{(a-b)(m_\phi-m_\psi)}{(a+b)^2} \rt] R(r)=K R(r).
\eea

\subsection{Extremal case}

Similar to 4D Kerr-Newman case, we need to use the linearly independent coordinates $\phi'=\a\phi+\b\psi$ and $\psi'=\g\phi+\d \psi$, with
\be \lt( \ba{cc} \a & \b \\ \g & \d \ea \rt) \in SL(2,\mb Z), \ee
to study the general pictures. The identification $e^{im_\phi\phi+im_\psi\psi}=e^{im_{\phi'}\phi'+im_{\psi'}\psi'}$ gives us
\be
m_\phi=\a m_{\phi'}+\g m_{\psi'},~~~m_\psi=\b m_{\phi'}+\d m_{\psi'}.
\ee
In order to get the picture corresponding to the $\phi'$ direction, we have to turn off the momentum mode along $\psi'$ direction by imposing $m_{\psi'}=0$. And then (\ref{z6}) becomes
\bea \label{z23}
&&\p_{r^2}(r^2-r_+^2)^2\p_{r^2}R(r)
    +\f{a b(a+b)^4}{4(r^2-r_+^2)^2}\lt(\o-\f{(\a+\b)m_{\phi'}}{a+b}\rt)^2 R(r)  \nn\\
&&+\f{(a+b)^4}{4(r^2-r_+^2)} \lt(\o-\f{(\a+\b)m_{\phi'}}{a+b}\rt)
                             \lt[ \o-\f{(\a-\b)(a-b)m_{\phi'}}{(a+b)^2} \rt] R(r)=K R(r).
\eea
As we have modified the definition of the
conformal coordinates slightly, the $r$ and $\p_r$ in the radial equation (\ref{z7}) becomes $r^2$ and $\p_{r^2}$ . If we further
set $m \ra m_{\phi'}$ in (\ref{z7}) and compare it with (\ref{z23}), then we find the agreement with the following identifications
\bea \label{z24}
&&\b_1^{\phi'}=-\f{(a+b)^2}{(\a-\b)(a-b)}\a_1^{\phi'},~~~
\g_1^{\phi'} =-\f{(\a b+\b a)\sr{ab}(a+b)^2}{(\a-\b)(a-b)}\a_1^{\phi'}, \nn\\
&&T_L^{\phi'}=\f{\sr{a b}}{\pi(\a b+\b a) },~~~n_L^{\phi'}=-\f{(\a+\b)\sr{ab}}{(\a b+\b a) (a+b)}.
\eea
Therefore from the study of the hidden conformal symmetry, we find the  same temperature as the one obtained in \cite{Chen:2011wm}.
According to (\ref{z12}) in the general picture the chiral CFT frequency $\o_L^\b$, charge $q_L^\b$ and chemical potential $\m_L^\b$ are respectively
\be\label{olq}
\o_L^{\phi'}=\f{(a+b)^2}{2(\a b+\b a)}\o,~~~
q_L^{\phi'}=m_{\phi'},~~~
\m_L^{\phi'}=\f{(\a-\b)(a-b)}{2(\a b+\b a)}.
\ee

The extremal 5D Kerr black hole entropy (\ref{z1}) can be written in terms of physical variables
\be \label{z20}
S_{BH}=2\pi\sr{J_\phi J_\psi}.
\ee
Taking variation on (\ref{z20}) directly and noting (\ref{z18}), we have
\be \label{z21}
\d S_{BH}=\f{\pi(b \d J_\phi+a\d J_\psi)}{\sr{a b}}.
\ee
And from the extremal constraint $M=(a+b)^2/2$, we have
\be
\d \ma M=\f{\d J_\phi+\d J_\psi}{a+b}.
\ee
Then we have (\ref{z21}) written as
\be \label{z22}
\d S_{BH}=\f{\pi(a+b)^2}{2\sr{ab}}\lt[ \d \ma M -\f{(a-b)(\d J_\phi-\d J_\psi)}{(a+b)^2}  \rt]
\ee
Identifying the relation (\ref{z22}) with that in dual CFT
\be
\d S_{CFT}^{\phi'}=\f{\d E_L^{\phi'}}{T_L^{\phi'}},
\ee
we get the conjugate charge
\be
\d E_L^{\phi'}=\f{(a+b)^2}{2(\a b+\b a)}\lt[ \d \ma M -\f{(a-b)(\d J_\phi-\d J_\psi)}{(a+b)^2} \rt].
\ee
The identifications of parameters are $\d \ma M=\o,~\d J_\phi=m_\phi=\a m_{\phi'}+\g m_{\psi'}$ and $\d J_\psi=m_\psi=\b m_{\phi'}+\d m_{\psi'}$. Now since  $m_{\phi'}=0$,  we have
\be
\td \o_L^\b=\d E_L^\b \lt( \d \ma M=\o;\d J_\phi=\a m_{\phi'};\d J_\psi=\b  m_{\phi'} \rt),
\ee
which is in accordance with (\ref{olq}).

Similarly, the $\psi'$ picture CFT dual can be confirmed in the same way.

\subsection{Non-extremal case}

For the general picture  we use the same coordinates $(\phi',\psi')$ as the extremal case and turn off the $\psi'$ direction momentum mode by setting $m_{\psi'}=0$, then we find that  Eq. (\ref{z5}) becomes
\bea \label{z13}
&&\p_{r^2}(r^2-r_+^2)(r^2-r_-^2)\p_{r^2}R(r)
   +\f{(r_+^2-r_-^2)\lt[\o-\lt(\a\O_+^\phi+\b \O_+^\psi\rt)m_{\phi'}\rt]^2}{4\k_+^2(r^2-r_+^2)}R(r) \nn\\
&& -\f{(r_+^2-r_-^2)\lt[\o-\lt(\a\O_-^\phi+\b \O_-^\psi\rt)m_{\phi'}\rt]^2}{4\k_-^2(r^2-r_-^2)}R(r)=K R(r),
\eea

Similarly if we set $r \ra r^2$, $m \ra m_{\phi'}$ in (\ref{z3}) and compare it with (\ref{z13}), then we have
\bea \label{z15}
&&T_L^{\phi'}=\f{r_+ + r_-}{2\pi (\a b+\b a)},~~~T_R^{\phi'}=\f{r_+ - r_-}{2\pi (\a b+\b a)},  \\
&&n_L^{\phi'}=-\f{(\a+\b)r_+^2(a+b)(r_+ + r_-)}{2(\a b+\b a)(r_+^2+a^2)(r_+^2+b^2)},\nn\\
&&n_R^{\phi'}=-\f{(\a-\b)r_+^2(a-b)(r_+ - r_-)}{2(\a b+\b a)(r_+^2+a^2)(r_+^2+b^2)},\nn
\eea
to find the agreement. Therefore we find the hidden conformal symmetry in the low frequency scattering off the 5D Kerr
black hole, which allows us to read the temperatures (\ref{z15}) of the dual CFT.

 With the central charge (\ref{z9}) and the temperatures (\ref{z15}) of the general picture we can reproduce successfully the Bekenstein-Hawking entropy (\ref{z2}) from the CFT entropy
\be
S^{\phi'}_{CFT}=\f{\pi^2}{3}(c_L^{\phi'} T_L^{\phi'}+c_R^{\phi'} T_R^{\phi'})=S_{BH}.
\ee

According to (\ref{z11}), in the general picture the CFT frequencies $(\o_L^{\phi'},\o_R^{\phi'})$, charges $(q_L^{\phi'},q_R^{\phi'})$ and chemical potential $(\m_L^{\phi'},\m_R^{\phi'})$ are respectively
\bea
&&\o_L^{\phi'}=\o_R^{\phi'}=\f{(r_+^2+a^2)(r_+^2+b^2)\o}{2(\a b+\b a)r_+^2},  \nn\\
&&q_L^{\phi'}=q_R^{\phi'}=m_{\phi'},  \nn\\
&&\m_L^{\phi'}=\f{(\a-\b)(a-b)}{2(\a b+\b a)},~~~\m_R^{\phi'}=\f{(\a+\b)(a+b)}{2(\a b+\b a)}.
\eea

From the first law of black hole thermodynamics
\be \label{z14}
\d S_{BH}=\f{\d \ma M-\O_+^\phi \d J_\phi-\O_+^\psi \d J_\psi}{T_H}
\ee
and its CFT dual
\be
\d S_{CFT}^{\phi'}=\f{\d E_L^{\phi'}}{T_L^{\phi'}}+\f{\d E_R^{\phi'}}{T_R^{\phi'}},
\ee
we get the conjugate charges $(\d E_L^{\phi'},\d E_R^{\phi'})$ in CFT as
\bea
&&\d E_L^\b=\f{(r_+^2+a^2)(r_+^2+b^2)\d \ma M-r_+^2(a-b)(\d J_\phi-\d J_\psi)}
              {2(\a b+\b a)r_+^2}, \nn\\
&&\d E_R^\b=\f{(r_+^2+a^2)(r_+^2+b^2)\d \ma M-r_+^2(a+b)(\d J_\phi+\d J_\psi)}{2(\a b+\b a)r_+^2}.
\eea
which could be identified with ($\td \o_L^\b, \td \o_R^\b$)
\bea
&&\td \o_L^{\phi'}=\d E_L^{\phi'} \lt( \d \ma M=\o;\d J_\phi=\a m_{\phi'};\d J_\psi=\b m_{\phi'} \rt), \nn\\
&&\td \o_R^{\phi'}=\d E_R^{\phi'} \lt( \d \ma M=\o;\d J_\phi=\a m_{\phi'};\d J_\psi=\b m_{\phi'} \rt).
\eea
Therefore, the hidden conformal symmetry indeed lead to a consistent dual CFT picture, supported by the low frequency scattering amplitude.

In the same way, we can get the $\psi'$-picture temperatures
\be T_L^{\psi'}=\f{r_+ + r_-}{2\pi (\g b+\d a)},~~~T_R^{\psi'}=\f{r_+ - r_-}{2\pi (\g b+\d a)}. \ee

\section{Conclusions and discussion}

In this paper we investigated the general hidden conformal symmetries of extremal and non-extremal 4D Kerr-Newman  and 5D Kerr black holes systematically. From the study of the hidden conformal symmetry of extreme black holes, we re-discovered the temperatures of the general dual CFT pictures suggested in \cite{Chen:2011wm}. Moreover, we showed that even for generic non-extremal black holes, there exist general hidden conformal symmetries in the low-frequency scattering off the black holes in certain region. The existence of such hidden conformal symmetries on the solution space suggesting that there are general CFT duals to these non-extremal black holes. Indeed, we read the temperatures from the identification of radial equation with the $SL(2,\mb R)$ Casimir and then reproduced the Bekenstein-Hawking entropies from microscopic CFT counting via the Cardy formula, under the assumption that the form of the central charges keep unchanged from the ones in extreme cases. We found further nontrivial support to the general CFT duals from the agreement of the real-time correlators with the CFT Euclidean correlators, where the first law of black hole thermodynamics is essential to determine the conjugate charges.

In this paper, we focused on the 4D Kerr-Newman black hole and 5D Kerr black hole, both of which have two $U(1)$ symmetries.  We believe for higher dimensional rotating black holes we can also have the similar picture and the general CFTs found in \cite{Chen:2011wm} can be generalized to non-extremal case, with the help of the hidden conformal symmetry.

\vspace*{10mm}

\noindent {\large{\bf Acknowledgments}}

The work was in part supported by NSFC Grant No. 10975005. BC would like to thank the
participants in ``The 2011 Workshop on String Theory and Cosmology" (Pusan, Korea) for
stimulating discussions and comments.

\vspace*{5mm}



\begin{thebibliography}{99}

\bibitem{Guica:2008mu}
  M.~Guica, T.~Hartman, W.~Song and A.~Strominger,
  ``The Kerr/CFT Correspondence,''
  Phys.\ Rev.\  D {\bf 80}, 124008 (2009)
  [arXiv:0809.4266 [hep-th]].

\bibitem{Bredberg:2011hp}
  I.~Bredberg, C.~Keeler, V.~Lysov and A.~Strominger,
  ``Cargese Lectures on the Kerr/CFT Correspondence,''
  arXiv:1103.2355 [hep-th].

\bibitem{Hartman:2008pb}
  T.~Hartman, K.~Murata, T.~Nishioka and A.~Strominger,
  ``CFT Duals for Extreme Black Holes,''
  JHEP {\bf 0904}, 019 (2009)
  [arXiv:0811.4393 [hep-th]].

\bibitem{Lu:2008jk}
  H.~Lu, J.~Mei and C.~N.~Pope,
  ``Kerr/CFT Correspondence in Diverse Dimensions,''
  JHEP {\bf 0904}, 054 (2009)
  [arXiv:0811.2225 [hep-th]].

\bibitem{Chen:2011wm}
  B.~Chen and J.~j.~Zhang,
  ``Novel CFT Duals for Extreme Black Holes,''
  arXiv:1106.4148 [hep-th].

\bibitem{Carlip:2011ax}
  S.~Carlip,
  ``Extremal and nonextremal Kerr/CFT correspondences,''
  JHEP {\bf 1104}, 076 (2011)
  [arXiv:1101.5136 [gr-qc]].

\bibitem{Matsuo:2009sj}
  Y.~Matsuo, T.~Tsukioka and C.~M.~Yoo,
  ``Another Realization of Kerr/CFT Correspondence,''
  Nucl.\ Phys.\  B {\bf 825}, 231 (2010)
  [arXiv:0907.0303 [hep-th]].\\
  ``Yet Another Realization of Kerr/CFT Correspondence,''
  Europhys.\ Lett.\  {\bf 89} (2010) 60001
  [arXiv:0907.4272 [hep-th]].

\bibitem{Castro:2009jf}
  A.~Castro and F.~Larsen,
  ``Near Extremal Kerr Entropy from AdS$_2$ Quantum Gravity,''
  JHEP {\bf 0912}, 037 (2009)
  [arXiv:0908.1121 [hep-th]].

\bibitem{Bredberg:2009pv}
  I.~Bredberg, T.~Hartman, W.~Song and A.~Strominger,
  ``Black Hole Superradiance From Kerr/CFT,''
  arXiv:0907.3477 [hep-th].


\bibitem{Hartman:2009nz}
  T.~Hartman, W.~Song and A.~Strominger,
  ``Holographic Derivation of Kerr-Newman Scattering Amplitudes for General
  Charge and Spin,''
  arXiv:0908.3909 [hep-th].

\bibitem{Cvetic:2009jn}
  M.~Cvetic and F.~Larsen,
  ``Greybody Factors and Charges in Kerr/CFT,''
  JHEP {\bf 0909}, 088 (2009)
  [arXiv:0908.1136 [hep-th]].

\bibitem{arXiv:1001.3208}
  B.~Chen and C.~-S.~Chu,
  ``Real-Time Correlators in Kerr/CFT Correspondence,''
  JHEP\ {\bf 1005}, 004  (2010)
  [arXiv:1001.3208 [hep-th]].

\bibitem{Becker:2010jj}
  M.~Becker, S.~Cremonini and W.~Schulgin,
  ``Extremal Three-point Correlators in Kerr/CFT,''
  arXiv:1004.1174 [hep-th].

\bibitem{Castro:2010fd}
  A.~Castro, A.~Maloney and A.~Strominger,
  ``Hidden Conformal Symmetry of the Kerr Black Hole,''
  Phys.\ Rev.\  D {\bf 82}, 024008 (2010)
  [arXiv:1004.0996 [hep-th]].

\bibitem{Chen:2010ik}
  B.~Chen and J.~Long,
  ``Hidden Conformal Symmetry and Quasi-normal Modes,''
  Phys.\ Rev.\  D {\bf 82}, 126013 (2010)
  [arXiv:1009.1010 [hep-th]].

\bibitem{Cvetic:2011hp}
  M.~Cvetic and F.~Larsen,
  ``Conformal Symmetry for General Black Holes,''
  arXiv:1106.3341 [hep-th].

\bibitem{Chen:2010as}
  C.~M.~Chen and J.~R.~Sun,
  ``Hidden Conformal Symmetry of the Reissner-Nordstr${\textrm{\"o}}$m Black Holes,''
  JHEP {\bf 1008}, 034 (2010)
  [arXiv:1004.3963 [hep-th]].

\bibitem{Chen:2010xu}
  B.~Chen and J.~Long,
  ``Real-time Correlators and Hidden Conformal Symmetry in Kerr/CFT
  Correspondence,''
  JHEP {\bf 1006}, 018 (2010)
  [arXiv:1004.5039 [hep-th]].

\bibitem{Chen:2010bh}
  B.~Chen and J.~Long,
  ``On Holographic description of the Kerr-Newman-AdS-dS black holes,''
  JHEP {\bf 1008}, 065 (2010)
  [arXiv:1006.0157 [hep-th]].


\bibitem{Chen:2010ywa}
  C.~M.~Chen, Y.~M.~Huang, J.~R.~Sun, M.~F.~Wu and S.~J.~Zou,
  ``Twofold Hidden Conformal Symmetries of the Kerr-Newman Black Hole,''
  Phys.\ Rev.\  D {\bf 82}, 066004 (2010)
  [arXiv:1006.4097 [hep-th]].

\bibitem{Krishnan:2010pv}
  C.~Krishnan,
  ``Hidden Conformal Symmetries of Five-Dimensional Black Holes,''
  JHEP {\bf 1007}, 039 (2010)
  [arXiv:1004.3537 [hep-th]].



\bibitem{Setare:2010sy}
  M.~R.~Setare and V.~Kamali,
  ``Hidden Conformal Symmetry of Rotating Black Holes in Minimal
  Five-Dimensional Gauged Supergravity,''
  Phys.\ Rev.\  D {\bf 82}, 086005 (2010)
  [arXiv:1008.1123 [hep-th]].


\bibitem{HiddenSymmetry}
  Y.~Q.~Wang and Y.~X.~Liu,
  JHEP {\bf 1008}, 087 (2010)
  [arXiv:1004.4661 [hep-th]].
  R.~Li, M.~F.~Li and J.~R.~Ren,
  Phys.\ Lett.\  B {\bf 691}, 249 (2010)
  [arXiv:1004.5335 [hep-th]].
  D.~Chen, P.~Wang and H.~Wu,
  arXiv:1005.1404 [gr-qc].
  M.~Becker, S.~Cremonini and W.~Schulgin,
  JHEP {\bf 1009}, 022 (2010)
  [arXiv:1005.3571 [hep-th]].
  H.~Wang, D.~Chen, B.~Mu and H.~Wu,
  arXiv:1006.0439 [gr-qc].
  Y.~Matsuo, T.~Tsukioka and C.~M.~Yoo,
  arXiv:1007.3634 [hep-th].
  K.~N.~Shao and Z.~Zhang,
  arXiv:1008.0585 [hep-th].
  M.~R.~Setare and V.~Kamali,
  arXiv:1008.1123 [hep-th].
  A.~M.~Ghezelbash, V.~Kamali and M.~R.~Setare,
  arXiv:1008.2189 [hep-th].
  B.~Chen, A.~M.~Ghezelbash, V.~Kamali and M.~R.~Setare,
  arXiv:1009.1497 [hep-th].
  R.~li and J.~R.~Ren,
  JHEP {\bf 1009}, 039 (2010), [arXiv:1009.3139 [hep-th]].
  J.~Rasmussen,
  J.\ Geom.\ Phys.\  {\bf 61}, 922 (2011)
  [arXiv:1009.4388 [gr-qc]].
  B.~Chen, C.~M.~Chen and B.~Ning,
  arXiv:1010.1379 [hep-th].
  M.~R.~Setare and V.~Kamali,
  JHEP {\bf 1010}, 074 (2010)
  [arXiv:1011.0809 [hep-th]]. C.~M.~Chen, V.~Kamali and M.~R.~Setare,
  Supergravity,''
  arXiv:1011.4556 [hep-th].
  B.~Chen and J.~j.~Zhang,
  Phys.\ Lett.\  B {\bf 699}, 204 (2011)
  [arXiv:1012.2219 [hep-th]].
  Y.~C.~Huang and F.~F.~Yuan,
  JHEP {\bf 1103}, 029 (2011)
  [arXiv:1012.5453 [hep-th]]. T.~Azeyanagi, N.~Ogawa and S.~Terashima,
  JHEP {\bf 1106}, 081 (2011)
  [arXiv:1102.3423 [hep-th]].
  H.~Zhang,
  JHEP {\bf 1103}, 009 (2011)
  [arXiv:1102.4721 [hep-th]].
  D.~A.~Lowe, I.~Messamah and A.~Skanata,
  arXiv:1105.2035 [hep-th].
  T.~Birkandan and M.~Cvetic,
  arXiv:1106.4329 [hep-th].


\bibitem{Myers:1986un}
  R.~C.~Myers and M.~J.~Perry,
  ``Black Holes in Higher Dimensional Space-Times,''
  Annals Phys.\  {\bf 172}, 304 (1986).


\bibitem{Cvetic:1997uw}
  M.~Cvetic and F.~Larsen,
  ``General rotating black holes in string theory: Grey body factors and event
  horizons,''
  Phys.\ Rev.\  D {\bf 56}, 4994 (1997)
  [arXiv:hep-th/9705192].




\bibitem{Chen:2010fr}
  B.~Chen, J.~Long and J.~j.~Zhang,
  ``Hidden Conformal Symmetry of Extremal Black Holes,''
  Phys.\ Rev.\  D {\bf 82}, 104017 (2010)
  [arXiv:1007.4269 [hep-th]].








\end{thebibliography}
\end{document}